\documentclass[aps,showpacs]{revtex4}
\usepackage{graphicx}
\begin{document}

\title{New measurement of cross section of evaporation residues from $^{\textrm{nat}}$Pr+$^{12}$C reaction: A comparative study on the production of $^{149}$Tb }

\author{Moumita Maiti\footnote{E-mail: moumita.maiti@saha.ac.in}}
\affiliation{Chemical Sciences Division, Saha Institute of Nuclear Physics, 1/AF, Bidhannagar, Kolkata-700064, India.}

\begin{abstract}

Production cross sections of evaporation residues, $^{149}$Tb, $^{150}$Tb, $^{151}$Tb and $^{149}$Gd, have been measured using the  stacked foil technique followed by off-line $\gamma$-spectrometry
in $^{12}$C induced reactions on naturally abundant mononuclidic praseodymium target in the 44-79 MeV incident energy range. Measured data have been interpreted comparing with previous measurements and theoretical prediction of nuclear reaction model code \textsc{PACE4}. About 5\% and 14\% of the theoretical cross sections have been measured for $^{149}$Tb and $^{150}$Tb, respectively.  The new cross sections of $^{149}$Tb complement those measured earlier by $\alpha$-spectrometry. Cross sections of $^{151}$Tb are comparable to the theory. Cumulative cross section of $^{149}$Gd sheds light on the nuclear reaction mechanism. In addition, a discussion has been made to show the feasibility of producing $^{149}$Tb in $p$- and $\alpha$-induced reactions on gadolinium isotopes.   

\end{abstract}


\maketitle

\section{Introduction}
\label{intro}

Due to the short range and high linear energy transfer in tissues, use of $\alpha$-emitting radionuclides has appeared promising over the $\beta$-emitters in targeted therapy, where localization of dose becomes important. It is therefore necessary to find more potential radionuclides to make the targeted $\alpha$-therapy a standard therapeutic modality. The rationale of selection of potential radionuclides depends on the nuclear property, methods of production and complexation behaviour with biological molecules. As a consequence, nuclear science has become a major part in developing this particular field.

So far, a few $\alpha$-emitters were found to be suitable for the targeted therapy. Due to short half-life and 3.97 MeV $\alpha$-particle, $^{149}$Tb (T$_{1/2}$=4.118 h, Decay mode: $\epsilon$ (83.3\%) and $\alpha$ (16.7\%)) is among the few promising $\alpha$-emitting radionuclides, which are projected for human clinical use \cite{allen1,McDevitt,allen2}. This gives impetus to investigate nuclear reactions in detail to produce $^{149}$Tb.

Production of proton rich $^{149}$Tb is possible only in accelerators by \\
(i) light ion ($p$, $d$, $\alpha$, $^{3}$He) induced reactions on Gd target \\
(ii) proton induced spallation reactions on high Z material, Ta or W, etc., and \\
(iii) heavy ion induced reactions. \\
However, hurdle lies in its production, particularly for applications, which demand significant quantity as well as purity of the radionuclide.  Only a few literatures are available that discussed on the production of $^{149}$Tb \cite{alex-simo,kossakowski,beyer}. 

\begin{figure} 
\begin{center}
\includegraphics [height=8.0cm]{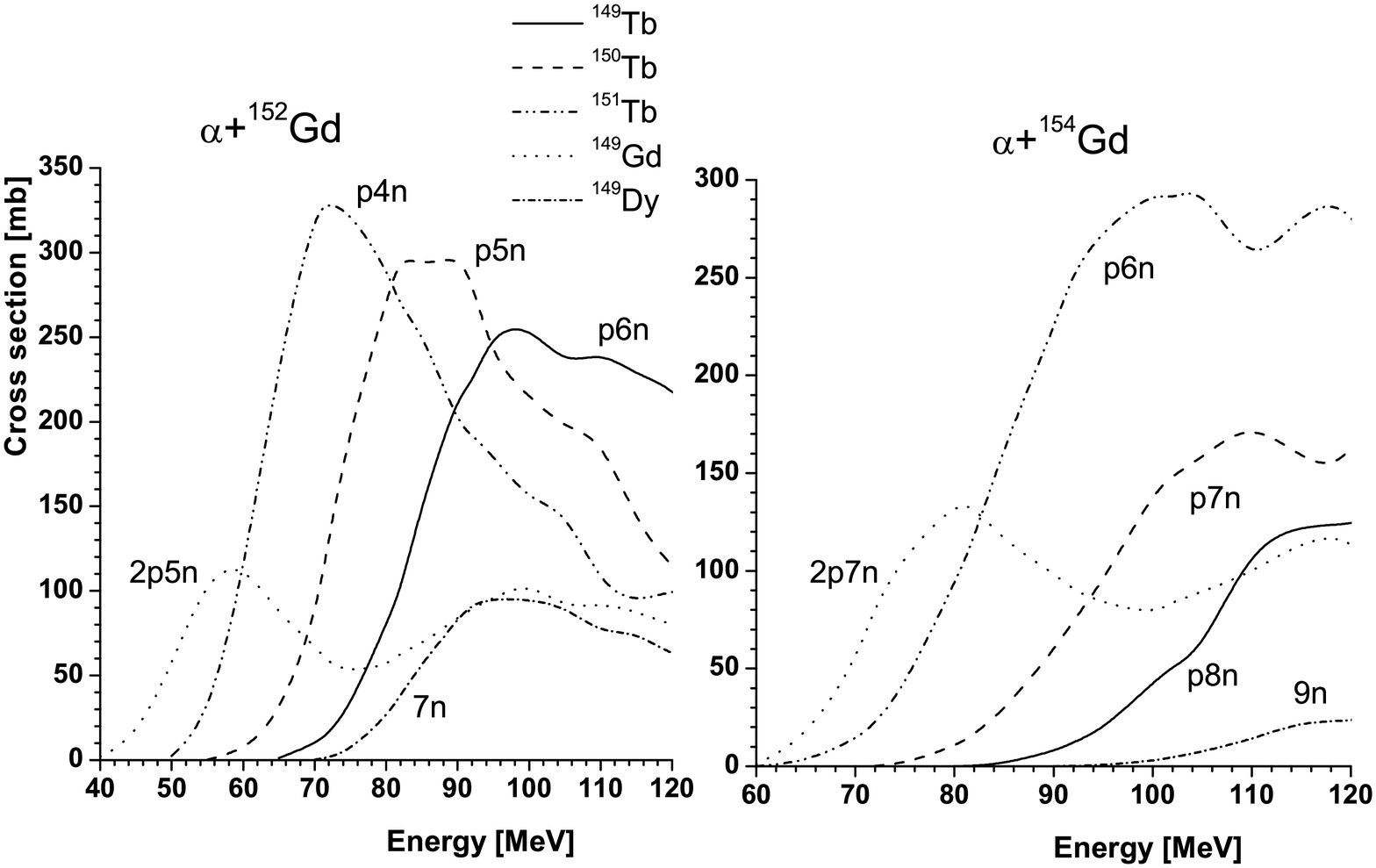}
\caption{Theoretical excitation functions of different reaction channels of $\alpha$+$^{152}$Gd and $\alpha$+$^{154}$Gd reactions calculated from \textsc{TALYS}.} 
\label{f5}
\end{center}
\end{figure}

Usually, light ion induced reactions are preferred to produce clinical radionuclides because of  high yield. However, in case of $^{149}$Tb, light ion induced productions suffer from a few shortcomings: (i) required projectile energy ($\approx$ 40-45 MeV protons (Fig. \ref{f4}) and 100 MeV $\alpha$-particles (Fig. \ref{f5})) is not available in common accelerators, (ii) suitable enriched Gd target may be required to maintain the purity of $^{149}$Tb, (iii) purification of $^{149}$Tb from the bulk target is not a easy task due to the similar chemical properties of lanthanides. A brief discussion on the light ion induced production of $^{149}$Tb has been appended in Sec. \ref{L3}. 

Due to the limited available facilities in the world, production of $^{149}$Tb from   1-2 GeV proton induced spallation reaction followed by online mass separation technique is also not feasible in practice. In this circumstance, it was assumed that heavy ion induced reactions may offer solution in producing $^{149}$Tb radionuclides. 

Several heavy ion reactions may lead to the production of $^{149}$Tb either directly or as a decay product of $^{149}$Dy. 
Alexander and Simonoff \cite{alex-simo} measured excitation functions  using $\alpha$-spectrometry from twelve heavy ion reactions that produce $^{149}$Tb from the deexcitation of the compound nucleus of Tb. Eight different heavy projectiles, $^{10}$B, $^{11}$B, $^{12}$C, $^{14}$N, $^{15}$N, $^{16}$O, $^{18}$O and $^{19}$F, were used in combination with a variety of target isotopes from Ba to Nd, among which $^{141}$Pr is the only naturally abundant mononuclidic target. It was found that the peak cross section measured in all target-projectile combinations is maximum 7\% of the theoretical estimation. This led to the conclusion that the compound nuclei having angular momentum less than 7.5$\pm$1.5$\hbar$ contributing to these reactions. It is also interesting to note that all the production routes offer comparable peak cross sections of $^{149}$Tb without any added advantage of using enriched targets.

Later Kossakowski et al. \cite{kossakowski} measured cross sections of evaporation residues from $^{12}$C- and $^{14}$N-induced reactions on $^{141}$Pr and five different enriched targets of Sm; $^{144}$Sm,$^{147}$Sm,$^{150}$Sm, $^{152}$Sm, and $^{154}$Sm, at 5-10 MeV/A incident energies using  $\gamma$-spectrometry with an aim to understand the competition between neutron, charged particle and $\gamma$-ray emission as a function of excitation energy, angular momentum, etc. The cross section data presented for $^{149}$Tb from $^{141}$Pr($^{12}$C,4n)$^{149}$Tb reaction were two orders of magnitude more with respect to \cite{alex-simo} and were likely for the high-spin isomeric state, $^{149}$Tb$^{m}$ (T$_{1/2}$=4.16 min). Among the various target-projectile combination studied in \cite{alex-simo,kossakowski}, $^{141}$Pr($^{12}$C,4n)$^{149}$Tb is the only common reaction found for the comparative study. 

No other experiment has been reported till now towards the production of $^{149}$Tb from this particular target-projectile combination  to justify the reported results of \cite{alex-simo,kossakowski} and the ambiguity, if any, lies between them.  However, we reported recently the yield of $^{149}$Tb and $^{149}$Gd under particular experimental condition and suggested the required chemistry for mutual separation of $^{149}$Tb and $^{149}$Gd  from praseodymium target \cite{mmTbchem}. 
$^{149}$Tb has dominating electron capture (83.3\%) decay mode and relatively long half-life compared to its high spin ($\frac{11}{2}^{-}$) isomeric state $^{149}$Tb$^{m}$, which directly decays to $^{149}$Gd (T$_{1/2}$=9.28 d) via electron capture (99.98\%). This situation favours off-line $\gamma$-spectrometric investigation of the production of $^{149}$Tb along with other residues. We therefore made an attempt to study the excitation functions from the $^{12}$C-induced reactions on natural $^{141}$Pr target in the 79-44 MeV incident energy range by off-line $\gamma$-spectrometry. The report provides an idea to visualize the production of impurity radionuclides along with the desired $^{149}$Tb. Reaction cross section data may also help to reduce certain impurities optimizing the reaction parameters. The definition of \textit{impurity} has been defined in view of nuclear medicine elsewhere in our recent article \cite{mmaiti}. We also report a comparative study on the production of $^{149}$Tb from $p$- and $\alpha$-particle induced reactions on Gd isotopes using the nuclear reaction model code \textsc{TALYS} \cite{talys}.

The experimental procedure and brief of the nuclear model calculations are presented in Sec. \ref{L1} and \ref{L2}, respectively. Section \ref{L3} discusses the results of the present study and Sec. \ref{L4} concludes the report.

\section{Experimental}
\label{L1}
The non-hygroscopic praseodymium oxide, Pr$_{6}$O$_{11}$ (Johnson, Matthey \& Co. Limited) was used as target material. The Pr$_{6}$O$_{11}$ targets of 2.5-3.0 mg/cm$^{2}$ thickness were prepared by centrifugation technique on aluminium foil backing of thickness 1.5 mg/cm$^{2}$. Target assembly was prepared by placing three such Pr$_{6}$O$_{11}$ targets each time and was bombarded by $^{12}$C-beams. Total six such target stacks were irradiated separately varying the incident energy with an overlap between them. The experiment was carried out at the BARC-TIFR Pelletron Accelerator facility, Mumbai, India. The integrated charge was recorded in each irradiation by an electron suppressed Faraday cup stationed at the back of the target assembly. Projectile energy at a target is the average of the incident and outgoing beam energies. Beam energy degradation in the target and the catcher foils have been calculated using the Stopping and Range of Ions in Matter (SRIM) \cite{srim}.

At the end of bombardment (EOB),  off-line $\gamma$-spectrometric study was carried out in each foil in a regular time interval up to seven days using an HPGe detector having resolution 2.13 keV at 1332 keV coupled with a PC based MCA. The background subtracted peak area count corresponding to a particular $\gamma$-ray energy is the measure of yield of an evaporation residue. The cross sections of the evaporation residues produced at various incident energies and from different reaction channels were calculated from the standard activation equation. A detail description of the calculation is available elsewhere \cite{mmTc}. The nuclear spectroscopic data used to calculate the production cross sections of the evaporation residue are enlisted in the Table \ref{t1} \cite{nudat2}. The $\gamma$-ray energies marked in bold have been used to determine the cross sections of the corresponding residue.


The uncertainties considered in the cross section measurement are :
(i)	calibration of detector $\approx$ 2\%	
(ii) target thickness in atoms/cm$^{2}$ $\approx$  5\% 
(iii) systematic error in beam current that was propagated to the cross section data $\approx$  10\%. 		
Uncertainty in incident beam energy may occur at the successive targets due to energy degradation in the aluminum catchers. However, according to ref \cite{wilken, kemmer}, the energy straggling is expected to be small even in case of lowest incident energy and hence was neglected in this work. In addition, error may occur in the measurement from the counting statistics. The associated error related to the cross section measurement was determined considering all the factors mentioned and the data presented up to 95\% confidence level. 

\section{Model calculation}
\label{L2}

\subsection{PACE4}

The fusion-evaporation code \textsc{PACE4} \cite{pace4}, modified version of  \textsc{PACE} (Projection Angular momentum Coupled Evaporation) \cite{pace}, working in the framework of \textsc
{LISE}++ \cite{lise} with several new feature, has been used to calculate the excitation function of residues expected to be produced in $^{12}$C-induced reaction on $^{{\textrm {nat}}}$Pr target. The deexcitation process of the excited nuclei has been calculated using the Hauser-Feshbach model. The transmission coefficients for light particle emission have been determined from the optical model potential with default optical model parameters. The code internally decides level densities and masses it needs during deexcitation. The Gilbert-Cameron level density prescription is used in the present work with $\textit{a}$, level density parameter, equals to A/12 MeV$^{-1}$. The ratio of $a_{f}$/$a_{n}$ is chosen as unity. Fission is considered as a decay mode. The finite range fission barrier of Sierk has been used.  The compound nuclear fusion cross section is determined by using the Bass method. The yrast parameter is taken as unity.

\subsection{TALYS}
The code \textsc{TALYS} \cite{talys} has been used to calculate the excitation functions of evaporation residues from $p$- and $\alpha$-particle induced reactions on $^{152}$Gd and $^{154}$Gd targets. It uses two-component exciton model to estimate preequilibrium emissions, Hauser-Feshbach formalism for equilibrium emissions and Coupled Channel analysis for direct reaction process. The level density formulation we have used is a combination of the constant temperature model by Gilbert and Cameron and the Fermi gas model. In this combination, total excitation energy range is divided into two regions: low and high energy region. Low energy region starts from 0 MeV to a certain energy, up to which constant temperature law is valid and the high energy part starts above that where Fermi gas model is used to calculate level densities.

\section{Results and discussion}
\label{L3}
\subsection{$^{12}$C-induced reactions}

The residual radionuclides produced in the target matrix from $^{12}$C+$^{141}$Pr reaction at different incident energies have been identified analysing the $\gamma$-spectra collected at different time intervals after EOB. Decay data clearly indicates the production of $^{149}$Tb, $^{150}$Tb, $^{151}$Tb and $^{149}$Gd radionuclides in the target matrix. Due to short half-life, no signature of the isomeric staes of $^{149}$Tb, $^{150}$Tb and $^{151}$Tb has been observed in the off-line $\gamma$-spectra. The nuclear reactions involved in producing the evaporation residues are tabulated in Table \ref{t1}. The measured cross sections of each residue are tabulated in Table \ref{t2}. Cross section data have been interpreted comparing with theoretical excitation functions of the residues estimated using the statistical model code \textsc{PACE4} \cite{pace4}. The cross sections of $^{149}$Tb and $^{150}$Tb are found to be maximum 5\% and 14\% of the theoretical estimations, respectively. 

\begin{figure} 
\begin{center}
\includegraphics [height=8.0cm]{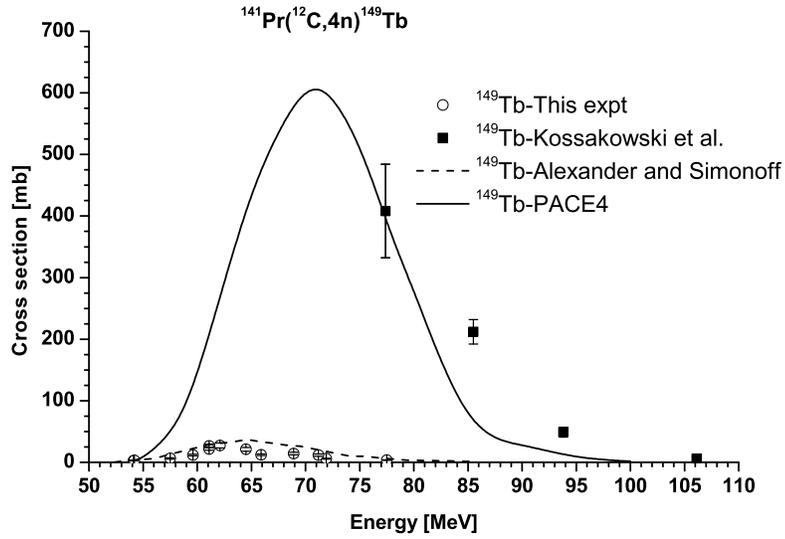}
\caption{Comparison of measured cross sections of $^{149}$Tb from $^{12}$C+$^{\textrm {nat}}$Pr with those measured by Alexander and Simonoff \cite{alex-simo} and Kossakowski et al. \cite{kossakowski} and calculated from \textsc{PACE4}.} 
\label{f1}
\end{center}
\end{figure}

Figure \ref{f1} compares measured cross sections of $^{149}$Tb from $^{12}$C+$^{141}$Pr reaction with those reported by Alexander and Simonoff \cite{alex-simo}, Kossakowski et al. \cite{kossakowski}, and the theoretical evaluation of \textsc{PACE4}. It is clear that the excitation function of $^{149}$Tb measured by off-line $\gamma$-spectrometry is comparable to that measured earlier using $\alpha$-spectrometry \cite{alex-simo}. The maximum cross section of 27.3$\pm$4.1 mb was obtained for $^{149}$Tb at 62.1 MeV, while it was maximum  of 36.7 mb at 64.6 MeV in the report of Alexander and Simonoff. It has also been observed that \textsc{PACE4} expects peak of the Gaussian at 72 MeV incident energy with $\approx$ 600 mb cross section for $^{149}$Tb while the measured excitation function peaks around 62 MeV incident energy with a cross section $<$5\% of the theoretical expectation. The cross sections of $^{141}$Pr($^{12}$C,4n) reaction channel measured by Kossakowski et al. by online $\gamma$-spectrometry are comparatively very high, 408 mb at 77.4 MeV incident energy, and do commensurate neither with our measurement nor with the measurement of Alexander and Simonoff \cite{alex-simo}. The high values were possibly the production cross sections of $^{149}$Tb$^{m}$, the high spin ($\frac{11}{2}^{-}$) isomer of $^{149}$Tb.  This observation indicates that the interaction of $^{12}$C-projectile with $^{141}$Pr target forms mostly excited compound nucleus $^{153}$Tb in the high spin state which preferentially decays to the short-lived high spin state $^{149}$Tb$^{m}$ (4.16 min, $\frac{11}{2}^{-}$). $^{149}$Tb$^{m}$ decays directly to $^{149}$Gd via $^{149}$Tb$^{m}$($\epsilon$/$\beta^{+}$)$^{149}$Gd reaction.  The $^{149}$Tb ($\frac{1}{2}^{+}$) is produced only from the low spin compound nuclei of $^{153}$Tb and this gives low cross section values for  $^{149}$Tb. Nevertheless, it is worthy to mention that two independent measurements of excitation function of $^{149}$Tb using $\alpha$- and $\gamma$-spectrometry complement each other with a confirmation of low cross section of $^{149}$Tb. 
 
\begin{figure} 
\begin{center}
\includegraphics [height=8.0cm]{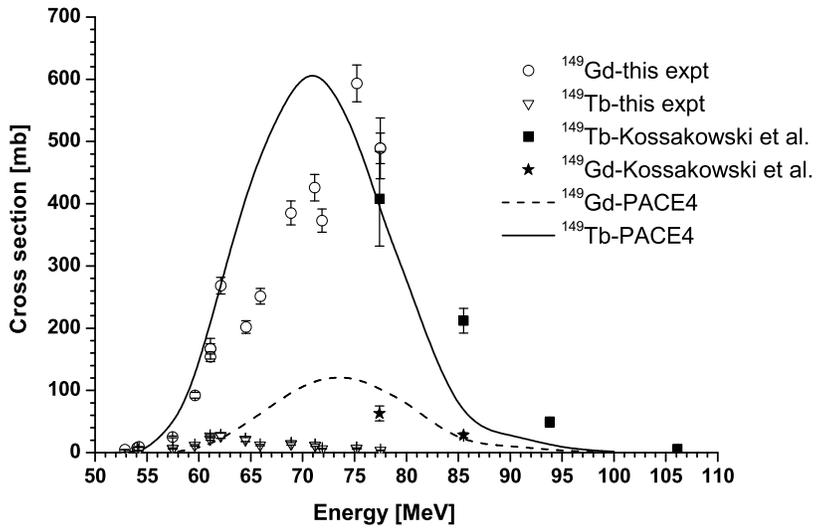}
\caption{Comparison of measured cross sections of $^{149}$Tb, $^{149}$Gd  from $^{12}$C+$^{\textrm {nat}}$Pr with those reported by Kossakowski et al. \cite{kossakowski} and theoretical calculation of \textsc{PACE4}.} 
\label{f2}
\end{center}
\end{figure}

Figure \ref{f2} compares production cross sections of $^{149}$Gd with the cross section values reported by Kossakowski et al. \cite{kossakowski} for $^{149}$Gd and $^{149}$Tb and the theoretical excitation functions of $^{149}$Gd and $^{149}$Tb estimated from \textsc{PACE4}. The $^{149}$Gd produced in the target matrix is probably the contribution from (i) direct reaction, $^{141}$Pr($^{12}$C,p3n)$^{149}$Gd (ii) indirect reactions, $^{141}$Pr($^{12}$C,4n)$^{149}$Tb$^{m}$($\epsilon$/$\beta^{+}$)$^{149}$Gd and $^{141}$Pr($^{12}$C,4n)$^{149}$Tb($\epsilon$/$\beta^{+}$)$^{149}$Gd. The cross section values we report here for $^{149}$Gd are the cumulative of direct and indirect productions. \textsc{PACE4} estimates production cross section of maximum 120 mb at 75 MeV incident energy for $^{149}$Gd in direct reaction. However, Kossakowski  et al. measured  63$\pm$12 and 28$\pm$4 mb cross section at 77.4 and 85.5 MeV respectively in $^{141}$Pr($^{12}$C,p3n)$^{149}$Gd reaction. Though only two experimental cross section values are available for $^{149}$Gd, they are  either comparable or lower than the theoretical expectation. The cumulative cross sections of $^{149}$Gd  more or less follow the theoretical excitation function of $^{141}$Pr($^{12}$C,4n)$^{149}$Tb with a little shift in peak energy to 75 MeV. The maximum cross section measured by off-line $\gamma$ spectrometry for $^{149}$Gd is as high as 593 mb at 75.2 MeV, the peak energy of the excitation function calculated for $^{149}$Gd from \textsc{PACE4}. Therefore, it may be concluded that the high cross section of $^{149}$Gd is due to the huge production of $^{149}$Tb$^{m}$, which essentially decays to $^{149}$Gd. The cross section values reported by Kossakowski et al. for $^{141}$Pr($^{12}$C,4n)$^{149}$Tb reaction nicely corroborate with the cumulative cross sections of $^{149}$Gd at the higher incident energies with an overlap of energy at 77.4 MeV.   At this particular energy, the cumulative cross section of $^{149}$Gd (489$\pm$24 mb) is nearly the sum of cross sections of $^{149}$Gd (63$\pm$12 mb) and $^{149}$Tb (408$\pm$76 mb) reported by Kossakowski et al. \cite{kossakowski}, whereas only 3.7$\pm$0.4 mb cross secton has been measured for low spin $^{149}$Tb. It certainly confirms that the production cross sections of $^{149}$Tb reported by Kossakowski et al. are the cross section of $^{149}$Tb$^{m}$. Analysis of the measured cross section data confirms that more than 85\% production of $^{149}$Gd comes from the decay of $^{149}$Tb$^{m}$. Production of $^{149}$Tb$^{m}$ dominates over the incident energy range, though the production of $^{149}$Tb is relatively higher below 65 MeV incident energy.   

\begin{figure} 
\begin{center}
\includegraphics [height=8.0cm]{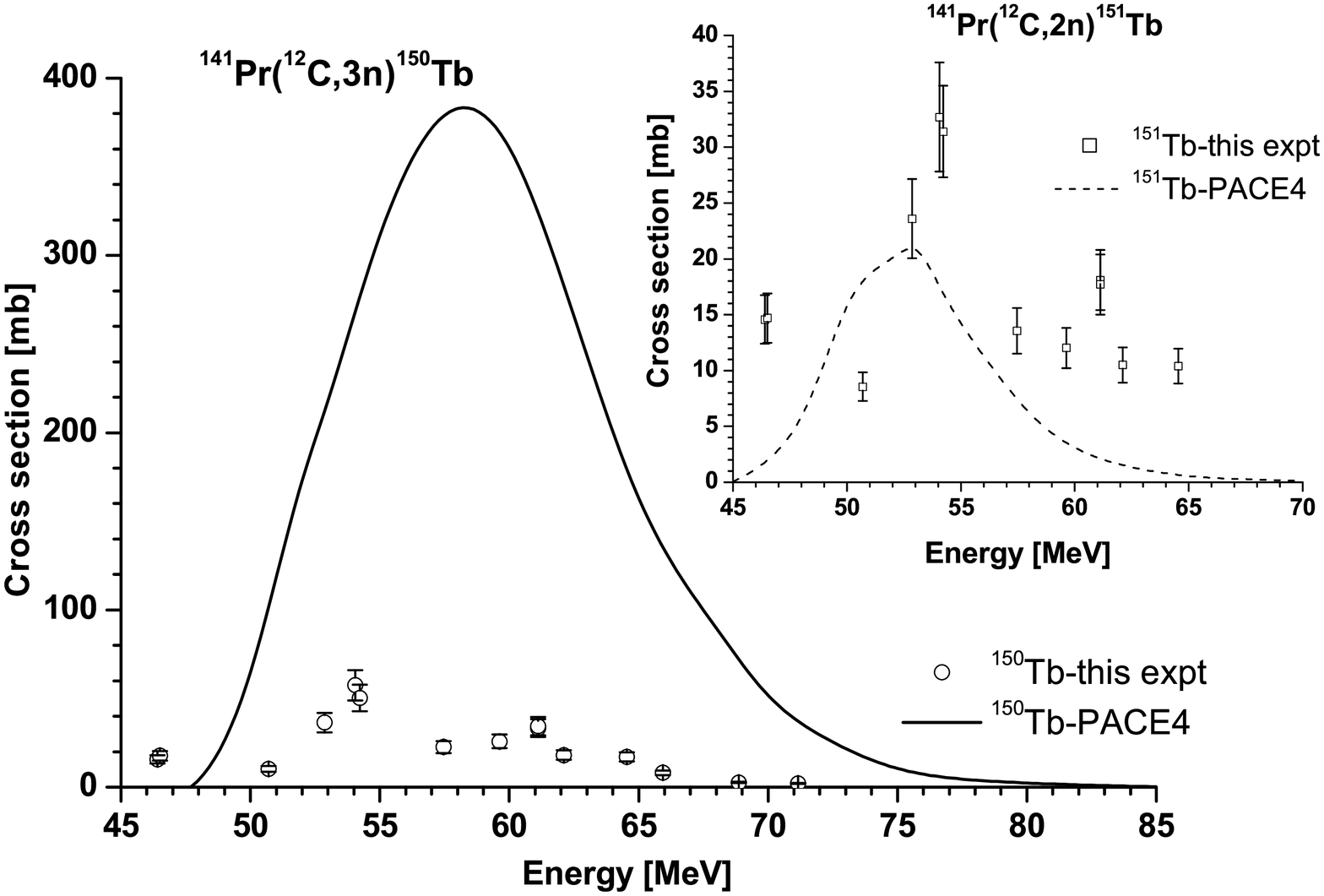}
\caption{Measured cross sections of $^{150}$Tb and $^{151}$Tb  from $^{12}$C+$^{\textrm {nat}}$Pr reaction have been compared with theoretical estimation of \textsc{PACE4}.} 
\label{f3}
\end{center}
\end{figure}

Figure \ref{f3} shows comparison of measured cross sections of $^{150}$Tb and $^{151}$Tb with those calculated from \textsc{PACE4}. The cross sections of $^{150}$Tb are found to be maximum 14\% of the theoretical estimation. The measured excitation function peaks at 54 MeV with an average 53.9$\pm$7.5 mb cross section for $^{150}$Tb. Kossakowski et al. measured 36$\pm$13 mb cross section of $^{150}$Tb at 77.4 MeV by online $\gamma$-spectrometry. However, we have observed no production of $^{150}$Tb in off-line $\gamma$-spectrometry above 72 MeV. Like $^{149}$Tb, low cross section of $^{150}$Tb may be due to the fact that the excited compound nucleus formed in high spin state prefers to produce high spin isomeric state $^{150}$Tb$^{m}$ (5.8 min, 9$^{+}$), which decays to long-lived $\alpha$-emitter $^{150}$Gd (1.8 My). The compound nuclei formed in the low spin state are only responsible in producing $^{150}$Tb (2$^{-}$). Measured cross sections of $^{151}$Tb were found comparable but absolute values are higher than that calculated from \textsc{PACE4}. In case of ($^{12}$C,2n) channel, the similar explanation is applicable towards production of $^{151}$Tb$^{m}$(25 s, $\frac{11}{2}^{-}$), but 93.4\% $^{151}$Tb$^{m}$ decays to $^{151}$Tb  via internal conversion. Therefore we have measured cross sections of $^{151}$Tb close to their expected value.

\subsection{Light ion induced reactions}
Gadolinium has seven naturally abundant isotopes. $p$- or $\alpha$-induced reactions on natural Gd will produce a variety of radionuclides and stable isotopes Tb, Dy, Gd etc., which are treated as impurity except the desired one. The reaction model code \textsc{ALICE} \cite{alice} is well established to investigate light ion induced reactions at low incident energies. However, \textsc{ALICE} does not consider direct reaction processes, which are prevalent in case of high energy $p$ or $\alpha$ projectiles. Therefore, we have estimated excitation functions of $p$ and $\alpha$-particle induced reactions on $^{152}$Gd and $^{154}$Gd targets, which are contributing to the production of $^{149}$Tb, using the nuclear reaction model code \textsc{TALYS} \cite{talys}. 

\begin{figure} 
\begin{center}
\includegraphics [height=8.0cm]{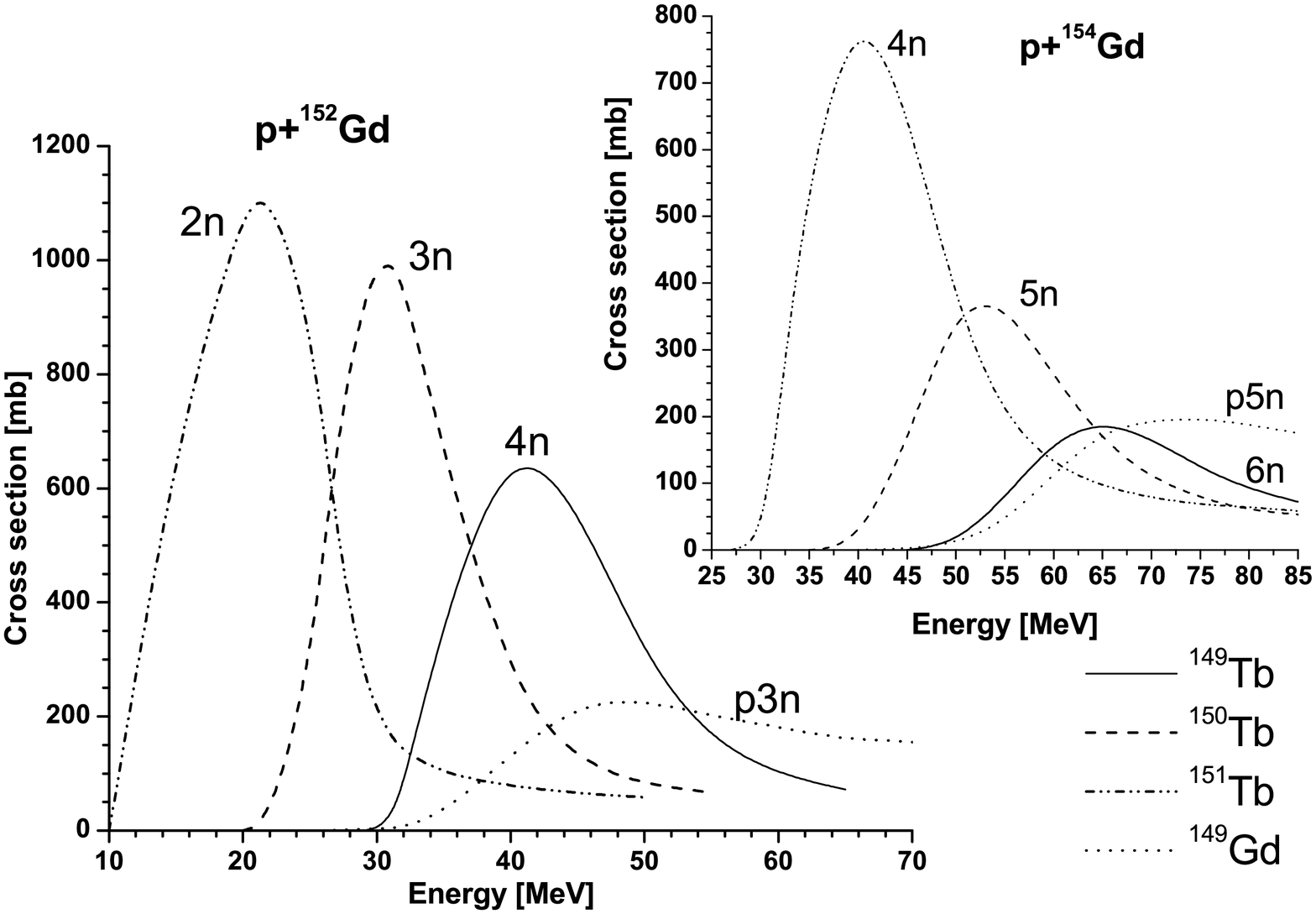}
\caption{Theoretical excitation functions of different reaction channels of $p$+$^{152}$Gd and $p$+$^{154}$Gd reactions calculated from \textsc{TALYS}.} 
\label{f4}
\end{center}
\end{figure}

Figure \ref{f4} shows excitation functions of $p$-induced reaction on $^{152}$Gd and $^{154}$Gd targets.  In $p$+$^{152}$Gd reaction, about 700 mb cross section is expected for $^{149}$Tb at 40 MeV incident energy, whereas only 200 mb cross section is obtained at 65 MeV from $p$+$^{154}$Gd reaction.  In view of the production of $^{149}$Tb, 40-45 MeV proton induced reaction on enriched $^{152}$Gd target is preferred. However, natural abundance of $^{152}$Gd is only 0.2\%, which certainly a disadvantage. Moreover, required high energy proton beam is not easy available. It is also observed from Fig. \ref{f4} that the possibility of production of impurity isotopes,  $^{150}$Tb, $^{151}$Tb and $^{149}$Gd along with $^{149}$Tb cannot be ignored irrespective of targets. 

Figure \ref{f5} shows the theoretical excitation functions of $\alpha$-particle induced reactions on $^{152}$Gd and $^{154}$Gd targets calculated from the code \textsc{TALYS}. In order to produce $^{149}$Tb, minimum 100 MeV $\alpha$-beam is required if enriched $^{152}$Gd target is used. About 250 mb cross section of $^{149}$Tb has been estimated in $^{152}$Gd($\alpha$, p6n)$^{149}$Tb reaction.  Decay of $^{149}$Dy produced via $^{152}$Gd($\alpha$, 7n) reaction is extected to enhance the production of $^{149}$Tb. However, comparable production of impurity ($^{150}$Tb, $^{151}$Tb, $^{149}$Gd) is also expected with $^{149}$Tb. It is not also reasonable to use $^{154}$Gd as production cross section of $^{149}$Tb is quite low (100 mb) and needs high energy $\alpha$-paricles (120 MeV). Therefore, production of pure $^{149}$Tb, located far away from the stability zone, is limited by the practical constraints even in light ion reactions.

\section{Conclusion}
\label{L4}
This article reports a new measurement of cross sections of evaporation residues produced in $^{12}$C-induced reaction on natural praseodymium target by off-line $\gamma$-spectrometry. Excitation function of the residues has been measured in 44-79 MeV energy range. Measured cross sections of $^{149}$Tb were found to be maximum 5\% of the theoretical estimation and in corroboration with those determined by Alexander and simonoff using $\alpha$-spectrometry. The high cumulative cross section measured for $^{149}$Gd satisfactorily explains the nuclear reaction phenomenon takes place during the production of $\alpha$-emitter $^{149}$Tb in the $^{12}$C+$^{141}$Pr reaction. It shows that more than 85\% of the cumulative cross section of $^{149}$Gd  comes from the decay of $^{149}$Tb$^{m}$. It also reports the first measurement of excitation functions of $^{150}$Tb $^{151}$Tb. 

A brief report presented on the $p$- and $\alpha$-induced production of $^{149}$Tb on Gd isotopes provides an idea of cross sections, required projectile energy, and target isotopes. It also reports the practical limitations involved in its production. Though there is enormous interest on the use of promissing $\alpha$-emitter $^{149}$Tb in targeted therapy, but its production is a genuine problem. None of the reactions reported here was found suitable for the production of $^{149}$Tb for clinical application. Therefore, further investigation is required to find other routes to produce $^{149}$Tb.

\begin{acknowledgments}

Author is indebted to Professor Susanta Lahiri for his encouragement and generous support to move on independent thinking. Thanks to \textsc{VECC} target laboratory for preparing Pr$_{6}$O$_{11}$ targets and the \textsc{TIFR} pelletron staff for their kind help during the experiments. Financial support from the Council of Scientific and Industrial Research (\textsc{CSIR}), India, is gratefully acknowledged. The work is a part of  the Saha Institute of Nuclear Physics-Department of Atomic Energy, \textsc{XI} five year plan project "Trace Analysis: Detection, Dynamics and Speciation (\textsc{TADDS})".

\end{acknowledgments}

\begin{table}
\caption{Nuclear spectrometric data of the radionuclides \cite{nudat2} produced through $^{12}$C+$^{\textrm {nat}}$Pr reactions.}
\label{t1}
\begin{tabular}{cccccc}
\hline
    Reaction & Product & Spin & T$_{1/2}$ & Decay mode & E$_{\gamma}$(keV) (I$_{\gamma}$ \%)  \\
\hline
   
 	 $^{141}$Pr($^{12}$C,4n) & $^{149}$Tb & $\frac{1}{2}^{+}$ & 4.118 h & $\epsilon$:83.3\%, $\alpha$:16.7\% & {\bf 164.98} (26.4), 352.24 (29.4)\\
 	 $^{141}$Pr($^{12}$C,4n) & $^{149}$Tb$^{m}$ & $\frac{11}{2}^{-}$ & 4.16 min & $\epsilon$:99.98\%, $\alpha$:0.02\% & 796.0 (97), 165.0 (7.3)\\
 	 $^{141}$Pr($^{12}$C,3n) & $^{150}$Tb & 2$^{-}$ & 3.48 h & $\epsilon$:100\%, $\alpha$:$<$0.05\% & {\bf 638.050} (72), 496.242 (14.6) \\
 	 $^{141}$Pr($^{12}$C,3n) & $^{150}$Tb$^{m}$ & 9$^{+}$ & 5.8 min & $\epsilon$:100\% & 638.050 (100), 650.36 (70) \\
 	 $^{141}$Pr($^{12}$C,2n) & $^{151}$Tb & $\frac{1}{2}^{+}$ & 17.609 h & $\epsilon$:99.99\%, $\alpha$:0.0095\% & 287.357 (28.3), {\bf 251.863} (26.3)\\
 	 $^{141}$Pr($^{12}$C,2n) & $^{151}$Tb$^{m}$ & $\frac{11}{2}^{-}$ & 25 s &  IT:93.4\%, $\epsilon$:6.6\% & 830.50 (3.3), 379.70 (6.3) \\
 	 $^{141}$Pr($^{12}$C,p3n), & $^{149}$Gd & $\frac{7}{2}^{-}$ & 9.28 d & $\epsilon$:100\%, $\alpha$:0.00043\% & 149.735 (48), {\bf 298.634} (28.6)  \\
   $^{149}$Tb$^{m}$($\epsilon$/$\beta^{+}$)& $^{149}$Gd & & & & \\
   
\hline
\end{tabular}
\end{table}

\begin{table}
\caption{Cross section of radionuclides produced in $^{12}$C+$^{\textrm {nat}}$Pr reactions.}
\label{t2}
\begin{tabular}{ccccc}
\hline
     					 E$_{lab}$ (MeV) & \multicolumn{4}{c} {\text{Cross section of isotopes (mb)}} \\\cline{2-5}
       				 								 & $^{149}$Tb & $^{150}$Tb & $^{151}$Tb & $^{149}$Gd \\
\hline
 							 61.1 & 26.6$\pm$2.7 & 33.3$\pm$5. & 18.1$\pm$2.7 & 166.9$\pm$8.3   \\ 
 	 						 54.1 & 2.5$\pm$0.2  & 57.5$\pm$8.1 & 32.7$\pm$4.9 & 7.5$\pm$0.4   \\
 	 						 46.4 &  -           & 15.7$\pm$2.3 & 14.6$\pm$2.2 & -  \\
\\
 	 						 77.5 & 3.7$\pm$0.4  & - & - & 489.1$\pm$24  \\
 	 						 71.9 & 6.1$\pm$0.6  & - & - & 373.0$\pm$18.7  \\
 	 						 65.9 & 12.2$\pm$1.2 & 8.1$\pm$1.2 & - & 251.4$\pm$12.6  \\
 							 59.6 & 11.9$\pm$1.2 & 25.8$\pm$3.9 & 12.0$\pm$1.8 & 92.1$\pm$4.6  \\
 	 						 52.9 &   -          & 36.4$\pm$5.5 & 23.6$\pm$3.5 & 4.7$\pm$0.2  \\
\\
 	 						 61.1 & 21.9$\pm$2.2 & 34.4$\pm$5.1 & 17.7$\pm$2.7 & 154.2$\pm$7.7  \\
 	 						 54.2 & 3.2$\pm$0.3  & 50.3$\pm$6.8 & 31.4$\pm$4.1 & 9.3$\pm$0.5  \\
 	 						 46.5 & -            & 17.7$\pm$2.7 & 14.7$\pm$2.2 & -  \\
\\	 						
 	 						 75.2 & 6.8$\pm$1.   & - & - & 593.3$\pm$29.  \\
 	 						 68.9 & 14.3$\pm$2.1 & 2.7$\pm$0.4 & - & 385.2$\pm$19.3  \\
 	 						 62.1 & 27.3$\pm$4.1 & 18.1$\pm$2.7 & 10.5$\pm$1.6 & 268.5$\pm$13.4  \\
\\	 					
 	 						 71.2 & 11.8$\pm$1.8 & -          	& - & 425.8$\pm$21.3  \\
 	 						 64.5 & 21.1$\pm$3.2 & 17.0$\pm$2.6 & 10.4$\pm$1.6 & 201.9$\pm$10.1  \\
 	 						 57.5 & 6.5$\pm$1.   & 22.6$\pm$3.4 & 13.6$\pm$2.0 & 25.3$\pm$1.3  \\
							 50.7 & -            & 10.3$\pm$1.5 & 8.6$\pm$1.3 & -  \\
\hline
\end{tabular}
\end{table}

\end{document}